\documentclass[english]{article}
\usepackage[T1]{fontenc}
\usepackage[latin9]{inputenc}
\usepackage{geometry}
\usepackage{color}
\usepackage{graphicx}
\usepackage{amsmath}
\usepackage{amssymb}
\usepackage{babel}


\usepackage{babel}

\begin{document}

\title{Probabilistic model of fault detection in quantum circuits}

\author{Anindita Banerjee and Anirban Pathak}

\maketitle
\begin{center}
Jaypee Institute of Information Technology University, Noida, India
\par\end{center}

 \begin{abstract}
 It is shown that the fault testing for quantum circuits does not follow conventional
classical techniques. If probabilistic gate like Hadamard gate is included in a circuit then the classical
notion of test vector is shown to fail. We have reported several new and distinguishing features of quantum
fault and also presented a general methodology for detection of functional faults in a quantum circuit. The
technique can generate test vectors for detection of different kinds of fault. Specific examples are given and
time complexity of the proposed quantum fault detection algorithm is reported.
\end{abstract}

\section{Introduction}

Since the introduction of quantum computation several protocols (such as quantum cryptography, quantum
algorithm, quantum teleportation) have established quantum computing as a superior future technology. Each of
these processes involves quantum circuits, which are prone to different kind of faults. Consequently it is
important to verify whether the circuit hardware is defective or not. The systematic procedure to do so is known
as fault testing. Normally testing is done by providing a set of valid input states and measuring the
corresponding output states and comparing the output states with the expected output states of the perfect
(fault less) circuit. This particular set of input vectors are known as test set~\cite{abap-nine}. If there
exist a fault then the next step would be to find the exact location and nature of the defect. This is known as
fault localization. A model that explains the logical or functional faults in the circuit is a fault model.
Conventional fault models include (i) stuck at faults, (ii) bridge faults and (iii) delay faults. These fault
models have been rigorously studied for conventional irreversible circuit. But with the advent of reversible
classical computing and quantum computing it has become important to enlarge the domain of the study on test
vectors. In the recent past people have realized this fact and have tried to provide good reversible fault
models \cite{abap-eight, abap-ten, abap-thirteen} which are independent of specific technology. The existing
reversible fault models are
\begin{enumerate}
\item{Single missing gate fault (SMF) where a single gate is missing in the circuit.} \item{Multiple missing
gate fault (MMGF) where many gates are missing in the circuit.} \item{Repeated gate fault (RGF) where same gate
is repeated consecutively many times.} \item{Partial missing gate fault (PGF) which can be understood as a
defective gate.} \item{Cross point fault  \cite{abap-thirteen} where the control points disappear from a gate or
unwanted control points appear on other gate.} \item{Stuck at fault model which includes single stuck at fault
(SSF) and multiple stuck at fault (MSF) for zero and one respectively.}
\end{enumerate}
These works are concentrated on circuits composed of gates from NCT, \footnote{This gate library has NOT, CNOT
and Toffoli gates. All these gates can also be achieved in the domain of reversible classical
computing.}%
and Generalized Toffoli gate libraries which are part of Maslov's benchmark \cite{abap-seven} but it all work in
the domain of classical reversible circuit. So most of the existing fault testing protocols \cite{abap-three,
abap-five, abap-eight, abap-eleven}, except \cite{abap-nine} have deterministic nature and are valid in the
domain of classical fault testing only. But advantage of quantum computing becomes prominent only when we use
superposition gates like Hadamard gate whereas hardly any effort has been made so far to include these gates in
the fault testing protocols. Further it is shown by Ito et al \cite{abap-twelve} that given a reversible circuit
C, it is NP-hard to generate a minimum complete test set for stuck at faults on the set of wires of C. These
facts have motivated us to aim to obtain an efficient algorithm for fault testing and generation of test set for
quantum circuits. Our effort in that direction yield several interesting characteristics of quantum fault. As
expected the distinguishing nature of quantum fault are only when we consider probabilistic gates and
superposition states in qubit line are considered. The fault testing models \cite{abap-eight, abap-ten,
abap-thirteen} studied so far are deterministic (D) but in quantum circuits containing superposition gates the
notion of determinism fails. To be precise if probabilistic gate like Hadamard gate is taken then the classical
notion of test vector fails. This has recently been realized by Perkowski \cite{abap-nine} and a new notion of
probabilistic test generation have been introduced by them \cite{abap-nine}. Present work follows independent
approach and reports several new and distinguishing features of quantum fault and provides a general methodology
for detection quantum fault.

To understand the basic nature of quantum fault, let us consider a Hadamard gate which maps
$\left|0\right\rangle $ to $\frac{1}{\sqrt{2}}\left|0\right\rangle +\frac{1}{\sqrt{2}}\left|1\right\rangle $ and
$\left|1\right\rangle $ to $\frac{1}{\sqrt{2}}\left|0\right\rangle -\frac{1}{\sqrt{2}}\left|1\right\rangle $.
Now if we consider $\left|x\right\rangle $ as test vector and get $\left|\bar{x}\right\rangle $ in the output
then we know that the gate exist but there is 50\% probability of getting $|x\rangle$ in the output and in that
case we shall not be able to conclude anything about the missing gate fault. Increase in number of trial will
increase the probability of detecting missing gate fault and after $n$ trials the probability of getting a
missing Hadamard fault will be $1-\frac{1}{2^{n}}.$ As the probability reduces to unity if an only if
$n=\infty$, therefore we can conclude that it is impossible to design a test vector which can always
deterministically identify a missing Hadamard gate. The conclusion remains valid for every superposition gates
and it is even valid if we do not work in the computational basis (i.e. even if you change the measurement
bases). In case of Hadamard gate $|0\rangle$ and $|1\rangle$are equally good test vectors but if we consider a 1
qubit gate $G_{1}=\left(\begin{array}{cc}
a_{11} & a_{21}\\
a_{31} & a_{41}\end{array}\right)$ in general then the probability of success of detecting a missing gate fault
after $n$ measurements by using $|0\rangle$ as test vector is $1-|a_{11}|^{2n}$ and the same with $|1\rangle$ as
test vector is $1-|a_{31}|^{2n}$. Consequently, both $|0\rangle$ and $|1\rangle$ can work as test vector but if
$|a_{11}|>|a_{31}|$ then $|1\rangle$ is a better test vector and if $|a_{31}|>|a_{11}|$ then $|0\rangle$ is a
better test vector. In general if we consider a generalized n qubit quantum gate $G_{n}$ which maps states
$|i\rangle$ (where $i$ varies from $0$ to $2^{n}-1$) as $G_{n}|i\rangle=\sum_{j=0}^{2^{n}-1}g_{ij}|j\rangle$
then we have to compare all $2^{n}-1$ values $g_{ii}$ and find out the lowest value among that. The state
$|i\rangle$ corresponding to the lowest value of $g_{ii}$ will provide the best probabilistic test vector.
Further we would like to note that in contrary to the
classical stuck at fault%
\footnote{which are of only two types, namely stuck at 0 and stuck at 1 } number of possible stuck at fault in
quantum circuit is infinite as a qubit line can stuck at $\alpha\left|0\right\rangle +\beta\left|1\right\rangle
$ $\forall$$\alpha,$$\beta,$$\epsilon\mathbb{C}$ : $\left|\alpha^{2}\right|+\left|\beta^{2}\right|=1$.
Practically, in a finite size circuit we do not need to consider all such stuck at faults but number of stuck at
fault models will not remain restricted to two.

In next section we have provided an algorithm to generate test vectors for quantum circuit. We have also
considered some specific circuits as examples to find different test vectors for the same and in the end we have
obtained the time complexity of the algorithm presented here. Section 3 is dedicated for conclusions.

\section{Methodology for detection of fault}
 The proposed fault testing algorithm can be logically divided into two parts. In first part we
find the total circuit matrix and in second part we find the Test vector for quantum circuit. It will comprise
of two qubit and one qubit gates and n qubit lines and is at present computable for m gates.
\begin{enumerate} \item{First let us consider an arbitrary quantum circuit having $n$ qubit lines and $m$ gates. The matrix of
each gate is expressed by a $2^{n}\times2^{n}$ matrix. This is easy because we just need to take tensor product
with identity operator for all those qubit lines which are not addressed by particular gate. For example, if we
consider a quantum circuit having three qubit lines and if the first gate (a NOT gate) is in second qubit line
then the matrix of this gate will be $I\otimes NOT\otimes I$. After this simple matrix product of all $m$
matrices (corresponding to $m$ gates) are obtained in sequence to obtain a $2^{n}\times2^{n}$ matrix which is
equivalent to the total circuit.} \item{In second part we find the matrices with all possible single and
multiple missing gate faults and relevant stuck at faults and compare them with the circuit matrix with no
faults. If any row of faulty circuit does not match with the resultant circuit matrix then the corresponding
input vector becomes the test vector. Test vector are not unique in case we obtain more than one test vector we
can follow the approach discussed in introduction in relation with the missing gate fault of a generalized $n$
qubit gate $G_{n}$. Further we would like to note that if a particular test vector appears for all faults then
it will comprise the test set. Otherwise, an optimized set of test vectors will comprise the test set. But in
this optimization process more importance should be given on the success rate than the order of the test set. To
be precise, one should choose all the test vectors which have highest possible success rate. In case two vectors
have same success rate (which is higher than all other possible test vectors) in detecting a fault then both are
initially kept in the list and then it is found how many other faults each of these vectors can detect. The one
which detects more faults get selected and become an element of the test set.}
\end{enumerate}

The algorithm discussed above can be clearly visualized through the following Fig. 1.

\begin{figure}[b]
\centering
\includegraphics[scale=.65]{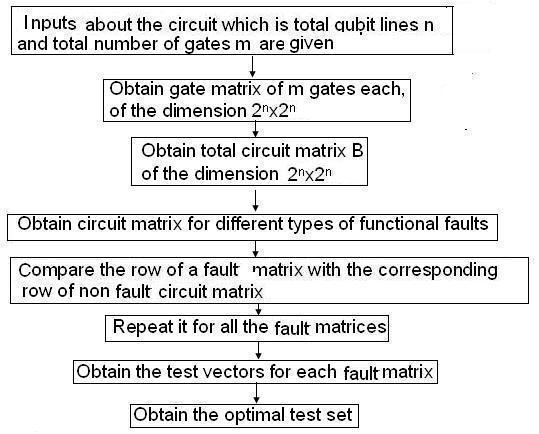}
\caption{Flowchart showing an algorithm for detection of fault in a quantum circuit} \label{Fig:1}
\end{figure}

\subsubsection{Specific examples}
\label{ABAPsubsec:1}
\begin{figure}[b]
\centering

\includegraphics[scale=.65]{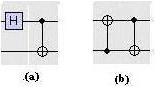}
\caption{Circuits for finding test vectors}\label{Fig:2}
\end{figure}

Consider an EPR circuit of 2 qubit lines consisting of a NOT gate in first qubit line and a CNOT gate with
target in second qubit line. The circuit is shown in Fig. 2(a). This circuit is a key component in teleportation
and all the other cases where entanglement generation is required. The total circuit matrix is given by B where

$B=B_{0}.$$B_{1}.B_{2}=\left(\begin{array}{cccc}
0 & 0 & 0 & 1\\
0 & 0 & 1 & 0\\
1 & 0 & 0 & 0\\
0 & 1 & 0 & 0\end{array}\right)$

$B_{0}$ is the Identity matrix

$B_{1}$ is the matrix of first gate tensor product with Identity

$B_{2}$ is the matrix of second gate

If first gate is missing, then the faulty circuit matrix will be equivalent to $B_{2}=\left(\begin{array}{cccc}
1 & 0 & 0 & 0\\
0 & 1 & 0 & 0\\
0 & 0 & 0 & 1\\
0 & 0 & 1 & 0\end{array}\right)$ and we note that all the rows of $B_{2}$ matrix and $B$ are different,
consequently all input vectors can detect the fault and if second gate is missing then the faulty circuit matrix
will be equivalent to $B_{1}=\left(\begin{array}{cccc}
0 & 0 & 1 & 0\\
0 & 0 & 0 & 1\\
1 & 0 & 0 & 0\\
0 & 1 & 0 & 0\end{array}\right)$ and we note that the last two rows are identical and thus
$\left|00\right\rangle $ and $\left|01\right\rangle $ can only detect the fault. Now if all the gates are
missing then again all the rows of $B_{0}=\left(\begin{array}{cccc}
1 & 0 & 0 & 0\\
0 & 1 & 0 & 0\\
0 & 0 & 1 & 0\\
0 & 0 & 0 & 1\end{array}\right)$ matrix and $B$ are not identical and all input vectors can detect the fault.
Thus we find that $\left|00\right\rangle $ and $\left|01\right\rangle $can detect all the missing gate faults.

Consider another circuit of 2 qubit line as showm in Fig. 2(b) consisting of a CNOT gate with target in second
qubit line and another CNOT gate with its target on first qubit line.

$B_{0}=I\otimes I$$=\left(\begin{array}{cccc}
1 & 0 & 0 & 0\\
0 & 1 & 0 & 0\\
0 & 0 & 1 & 0\\
0 & 0 & 0 & 1\end{array}\right)$, $B_{1}=\left(\begin{array}{cccc}
1 & 0 & 0 & 0\\
0 & 1 & 0 & 0\\
0 & 0 & 0 & 1\\
0 & 0 & 1 & 0\end{array}\right)$, $B_{2}=\left(\begin{array}{cccc}
1 & 0 & 0 & 0\\
0 & 0 & 0 & 1\\
0 & 0 & 1 & 0\\
0 & 1 & 0 & 0\end{array}\right)$

Thus total circuit matrix

$B=B_{0}.$$B_{1}.B_{2}=\left(\begin{array}{cccc}
1 & 0 & 0 & 0\\
0 & 0 & 0 & 1\\
0 & 1 & 0 & 0\\
0 & 0 & 1 & 0\end{array}\right)$

If first gate is missing, then the faulty circuit matrix will be equivalent to $B_{2}$ and we note that first
and second rows of $B_{2}$ matrix and $B$ are identical and thus $\left|10\right\rangle $ and
$\left|11\right\rangle $ input vectors can detect the fault and if second gate is missing then the faulty
circuit matrix will be equivalent to $B_{1}$ and we note that the second and third rows are identical and thus
$\left|01\right\rangle $ and $\left|10\right\rangle $ can detect the fault. Now if all the gates are missing
then as we can see that only first row is identical with circuit matrix and thus $\left|01\right\rangle ,$
$\left|10\right\rangle $ and $\left|11\right\rangle $ input vectors can detect the fault. Thus we find that
$\left|10\right\rangle $ can detect all the missing gate faults. A set of other examples for single missing gate
fault (SMF) and multiple missing gate fault (MMF) with their test vectors and the nature of fault which is
probabilistic (P) or deterministic (D) is \textcolor{black}{given in Table I. For large circuits as in
 \cite{abap-four}} we divide it into smaller sub circuits and find test vectors for each sub circuit.
This is shown in Table II.
\begin{figure}[h]
\centering
\includegraphics[scale=.9]{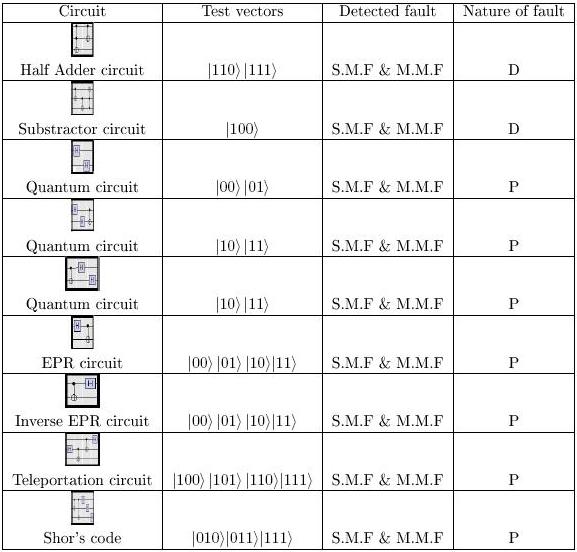}
\caption{Table I: Test vectors for different quantum circuit}
\label{Fig:3}       
\end{figure}

\begin{figure}[h]
\centering
\includegraphics[scale=.65]{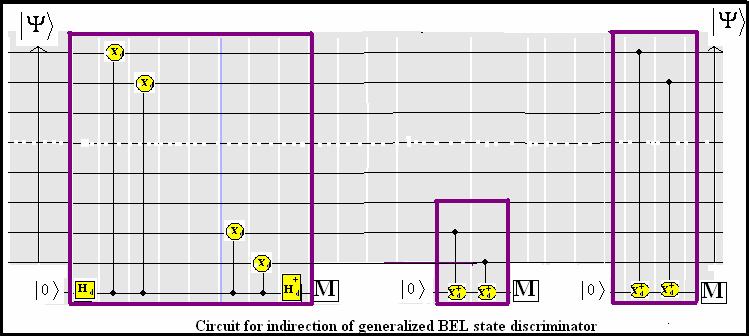}
\caption{Circuit for non-destructive generalized orthonormal qudit Bell state discriminator}
\label{Fig:4}       
\end{figure}

\begin{figure}[h]

\includegraphics[scale=.65]{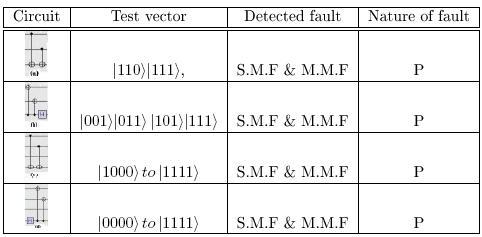}
\caption{Table II: Test vectors for non-destructive generalized orthonormal qudit Bell state discriminator
circuit.}
\label{Fig:5}       
\end{figure}
\subsection{Time complexity of the fault detection algorithm}

The first step require $m2^{2n}$ multiplications to obtain m matrices of $2^{n}\times2^{n}$ dimension which
corresponds to $m$ gates. The next step to obtain the resultant matrix require $(m-1)2^{3n}$ multiplications and
$(m-1)2^{2n}$ additions. Thus it requires $\mathcal{O}\left(m2^{3n}\right)$ steps to obtain the equivalent
matrix of the circuit similarly, it requires the same number of steps to find fault matrices corresponding to
each fault. So if we wish to check $p$ faults it requires $\mathcal{O}\left(pm2^{3n}\right)$ steps to construct
all the matrices. Now in order to compare the matrices we require $p\times2^{2n}$ steps in the worst case. Thus
the total time complexity is $\mathcal{O}\left(pm2^{3n}\right)$. As it has a linear relation with the number of
fault to be considered and the total number of stuck at fault is infinite so we can not detect all faults in
finite time but the methodology will work for all practical purposes where the number of faults of practical
interest is finite because of the physical restrictions.

\section{Conclusions}

We have followed an independent approach for generation of test set for quantum circuits and have reported
several new and distinguishing features of quantum fault. We have seen that for a quantum gate the classical
notion of test vector fails and theoretically it is impossible to determine a test set for hadamard gate.
Further we have observed that in contrary to the classical stuck at fault the number of possible stuck at fault,
in quantum circuit is infinite as the qubit line can be stuck at $\alpha\left|0\right\rangle
+\beta\left|1\right\rangle $ $\forall$$\alpha,$$\beta,$$\epsilon\mathbb{C}$ :
$\left|\alpha^{2}\right|+\left|\beta^{2}\right|=1$. It is also observed that the test set for quantum circuit,
for stuck at fault, is different from that of missing gate fault. For an odd number of repeated gate fault for
an optimized circuit is equivalent to a missing gate fault. In case of an even number it will not be detected.
It has been shown that the quantum faults are infinite in number and many of them cannot be detected
deterministically. Above observations suggested that the systematic procedure for generation of quantum test set
is would be different from the classical procedure. A methodology of generation of test set for quantum circuit
is prescribed here. Only few simple examples have been discussed here but since the algorithm is robust and
valid for any quantum or reversible circuits, following the same methodology, in future test vectors for other
useful circuits (existing and new) can be presented. Further, attempt to reduce the complexity of the method can
be made in future.


\end{document}